\documentclass[aip,rsi,amsmath,amssymb,reprint]{revtex4-1}

\usepackage{graphicx}
\usepackage{dcolumn}
\usepackage{color}
\usepackage{bm}
\usepackage{ulem}

\usepackage[breaklinks]{hyperref}
\usepackage{hyperref}
\usepackage{breakurl}
\usepackage{url}
\usepackage[all]{hypcap}

\begin{document}


\title{40-tesla pulsed-field cryomagnet for single crystal neutron diffraction\\}

\author{F. Duc}
\affiliation{Laboratoire National des Champs Magn\'etiques
Intenses, CNRS-INSA-UGA-UPS, F-31400 Toulouse, France}

\author{X. Tonon}
\affiliation{Institut Laue-Langevin, F-38000 Grenoble, France}

\author{J. Billette}
\affiliation{Laboratoire National des Champs Magn\'etiques
Intenses, CNRS-INSA-UGA-UPS, F-31400 Toulouse, France}

\author{B. Rollet}
\affiliation{Institut Laue-Langevin, F-38000 Grenoble, France}

\author{W. Knafo}
\affiliation{Laboratoire National des Champs Magn\'etiques
Intenses, CNRS-INSA-UGA-UPS, F-31400 Toulouse, France}

\author{F. Bourdarot}
\affiliation{Service de Mod\'{e}lisation et d'Exploration des Mat\'{e}riaux,
Universit\'{e} Grenoble Alpes et Commissariat \`{a} l'Energie Atomique, INAC, 38054 Grenoble, France}

\author{J. B\'{e}ard}
\affiliation{Laboratoire National des Champs Magn\'etiques
Intenses, CNRS-INSA-UGA-UPS, F-31400 Toulouse, France}

\author{F. Mantegazza}

\author{B. Longuet}
\affiliation{Service de Mod\'{e}lisation et d'Exploration des Mat\'{e}riaux,
Universit\'{e} Grenoble Alpes et Commissariat \`{a} l'Energie Atomique, INAC, 38054 Grenoble, France}

\author{J.E. Lorenzo}
\affiliation{Institut N\'{e}el, CNRS, Bo\^{\i}te Postale 166X, F–38043 Grenoble Cedex, France}

\author{E. Leli\`{e}vre-Berna}
\affiliation{Institut Laue-Langevin, F-38000 Grenoble, France}

\author{P. Frings}
\affiliation{Laboratoire National des Champs Magn\'etiques
Intenses, CNRS-INSA-UGA-UPS, F-31400 Toulouse, France}

\author{L.-P. Regnault}
\affiliation{Institut Laue-Langevin, F-38000 Grenoble, France}

\date{\today}

\begin{abstract}
We present the first long-duration and high duty cycle $40$-tesla pulsed-field cryomagnet
addressed to single crystal neutron diffraction experiments at temperatures down to 2\,K. The magnet produces
a horizontal field in a bi-conical geometry, $\pm$15 and  $\pm$30$^\circ$ upstream and
downstream of the sample, respectively. Using a 1.15\,MJ mobile generator, magnetic field pulses of 100\,ms
length are generated in the magnet, with a rise time of 23\,ms and a repetition rate of
6-7 pulses per hour at 40\,T. The setup was validated for neutron diffraction on the
CEA-CRG three-axis spectrometer IN22 at the ILL.

\end{abstract}

\pacs{Valid PACS appear here}
\maketitle

\section{\label{sec:level1}Introduction}
Limited for a long time to 15\,T and finally to 17\,T produced by superconducting coils \cite{Prokes2001,Holmes2012},
the magnetic field strength of DC magnets for neutron experiments has been raised
very recently to 26\,T at the Helmholtz-Zentrum Berlin (HZB) \cite{Smeibidl2010,Smeibidl2016}.
A new series-connected hybrid magnet with a horizontal field
orientation was designed and constructed by the HZB and
the National High Magnetic Field Laboratory (NHMFL), Tallahassee (Florida, USA)\cite{Steiner2006}.
Combined with the Extreme Environment Diffractometer (EXED) \cite{Prokhnenko2015},
the High Field Magnet (HFM) enables neutron studies in static fields up to 26\,T
and temperatures down to 0.65\,K \cite{Prokhnenko2017,Prokes2017,comment}.

Encouraged by the development of x-ray diffraction and spectroscopy in pulsed field up to 40\,T
at synchrotron sources \cite{Narumi2006, Matsuda2006, Frings2006, Linden2008, Islam2009},
significant activity aiming at increasing the magnetic field available at neutron facilities
has arisen all over the world.
Whereas the duty cycle \cite{duty} is rather low in pulsed field set-up, the combination of pulsed field and neutron
scattering environments is an alternative choice, of moderate cost,
whose relevance has been proven over the past decade.
Several pulsed field devices using mini-coil systems developed
by the Nojiri group from the Institute for Materials Research (Tohoku University, Japan) have been used
at different neutron facilities: a 30\,T solenoid magnet with pulse durations
of the order of 1 to 10 ms and a 25\,kJ capacitor bank have been first installed
in the research reactor JRR3 (JAEA, Tokai, Japan) \cite{Ohoyama2007}
and later at the japanese spallation neutron sources J-PARC \cite{Maekawa2009, Ohoyama2011} and at the Spallation Neutron Source
(SNS), Oak Ridge National Laboratory (USA) \cite{Nojiri2011}.
Currently, a 40-tesla horizontal coil system and a 250\,kJ compact power supply are being
developed by the same group and planned to be used at J-PARC
and at ISIS in United Kingdom \cite{Stone2014}.
Meanwhile, two successive experiments performed on the IN22 spectrometer from the Commissariat \`{a} l'Energie Atomi\-que (CEA)
at the high-flux neutron source of the Institut Laue Langevin (ILL, Grenoble) have demonstrated the feasibility of the technique:
combining a mini-coil system of the Nojiri group and a transportable
pulsed field generator \cite{Frings2006} developed by the Laboratoire National des
Champs Magn\'etiques Intenses (LNCMI, Toulouse), they
led to the first successful experiments associating
pulsed magnetic fields up to 30\,T and neutron diffraction \cite{Yoshii2009, Matsuda2010}.
These experiments have also clearly demonstrated the necessity of improving the duty cycle
of the coils, in particular by increasing the duration of the field pulse by
at least an order of magnitude.

In this paper, we present the first long-duration and high duty cycle pulsed-field
cryomagnet producing a maximum horizontal field of 40\,T designed for neutron diffraction.
Compared to mini-coils, this magnet offers improved experimental conditions for single-crystal neutron diffraction studies.
Remarkably better counting statistics with fewer number of pulses, i.e. a ten to fifteen times superior duty cycle,
result from an increased pulse duration combined with a faster cooling of the magnet.
A twice as wide angular access in the scattering plane allows probing larger parts of the reciprocal space.
In addition, longer rising times greatly reduce heating by eddy currents so that lower temperatures and accurate control of the sample temperature can be achieved.
Finally, the larger bore provides not only a larger volume for the sample
but also allows for a better and more sophisticated sample environment than the mini-coil system.
A sapphire cold finger connected to a heat exchanger inside the cone of the magnet allows to cool the sample down to 2\,K
whilst decoupling mechanically and thermally the sample from the magnet.
\begin{figure*}
\includegraphics[width=1.7\columnwidth]{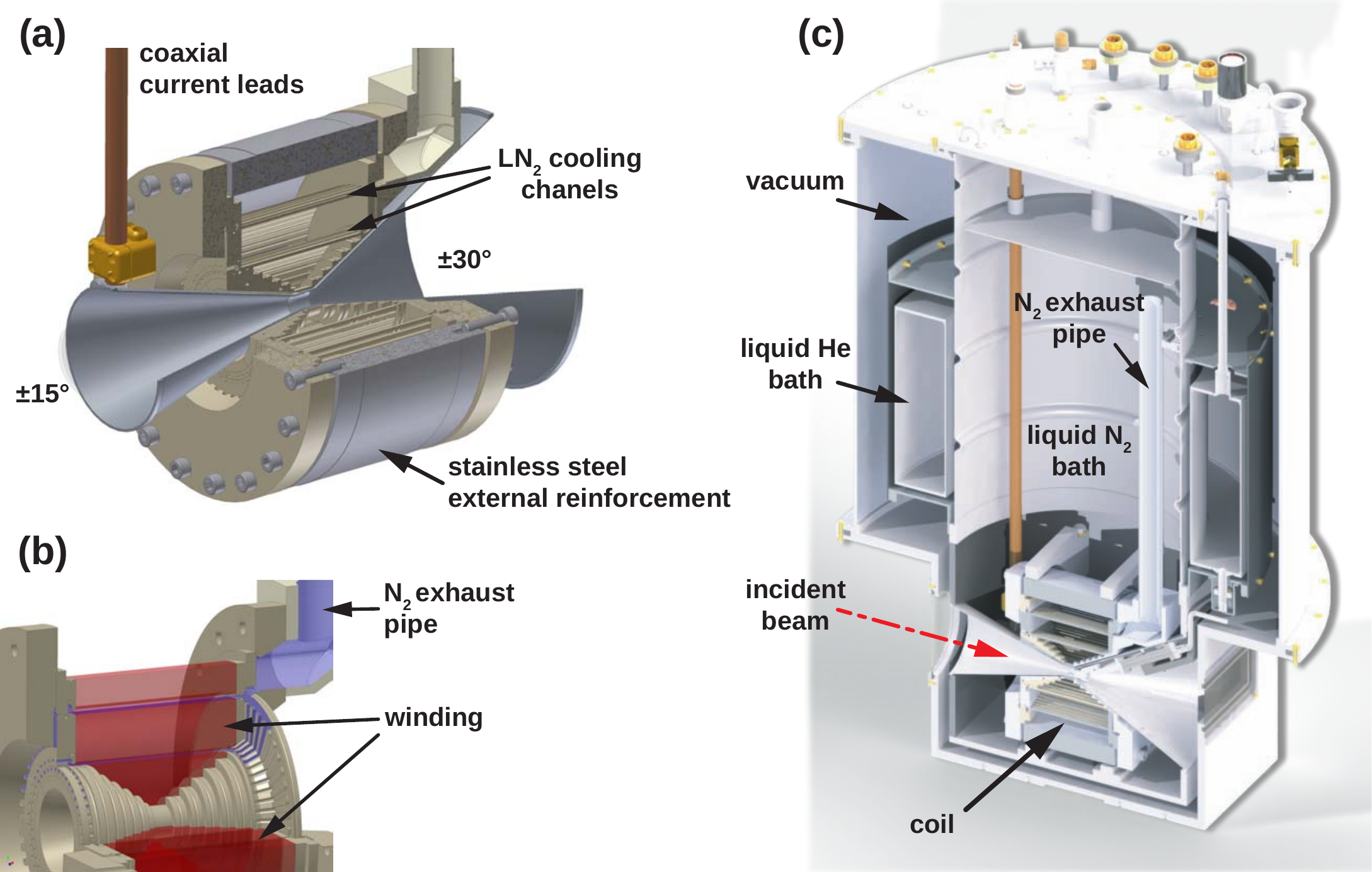}
\caption{Three-dimensional views of the magnet for neutron diffraction.
(a) The coil is wound around a stainless steel double-cone. The nitrogen circulates through
cooling channels and exits through the exhaust pipe highlighted in blue colour in view (b).
(c) Overview of the 40-tesla cryomagnet showing the coil at the bottom of the
liquid nitrogen bath and the liquid He jacket used to cool the sample down to 2\,K.
}
\label{coil_MAGFINS}
\end{figure*}
\section{Cryomagnet design and performances}
The coil has been designed and built by the LNCMI-Toulouse. Its construction is based
on the combination of a ``standard'' rapid-cooling technique \cite{Frings2008} with a winding around
a conical bore, as was already performed for a coil developed for x-ray scattering
\cite{Billette2012}. The rapid-cooling technique consists in inserting liquid nitrogen cooling channels
directly into the winding to increase the cooling efficiency of the coil and reach
a high duty cycle. The major technological novelty is the winding of the coil around
a double-cone constituting a main part of the cryostat since the cryomagnet should
have conical openings on two sides for maximising neutron beam access.
For that purpose, a highly non-magnetic 1\,mm thick double-cone of stainless steel
X2CrNi19-11 (EN 1.4306 / AISI 304L)
was machined to provide opening angles of $\pm 15^\circ$ and $\pm 30^\circ$
in the scattering plane for respectively the incident and scattered beams
(see Fig.~\ref{coil_MAGFINS}).

The coil was wound of CuAg wire (CuAg 0.08\%; ultimate tensile strength of 384\,MPa at room temperature
and 515\,MPa at 77\,K) reinforced with Zylon using the optimized reinforcement density technique \cite{Bockstal1991}.
It is composed of 32 layers of 4 to 40 turns. The inner diameter
(winding diameter) is $\O 24.5$\,mm ($\O 16.5$\,mm inside the stainless steel double-cone).
Its external diameter and height are $\O280$\,mm and 210\,mm, respectively.
To ensure high mechanical strengths, the wire was wound onto a coil-body of glass
fiber-epoxy composite (FR4/G11).

\begin{figure}
\includegraphics[width=\columnwidth]{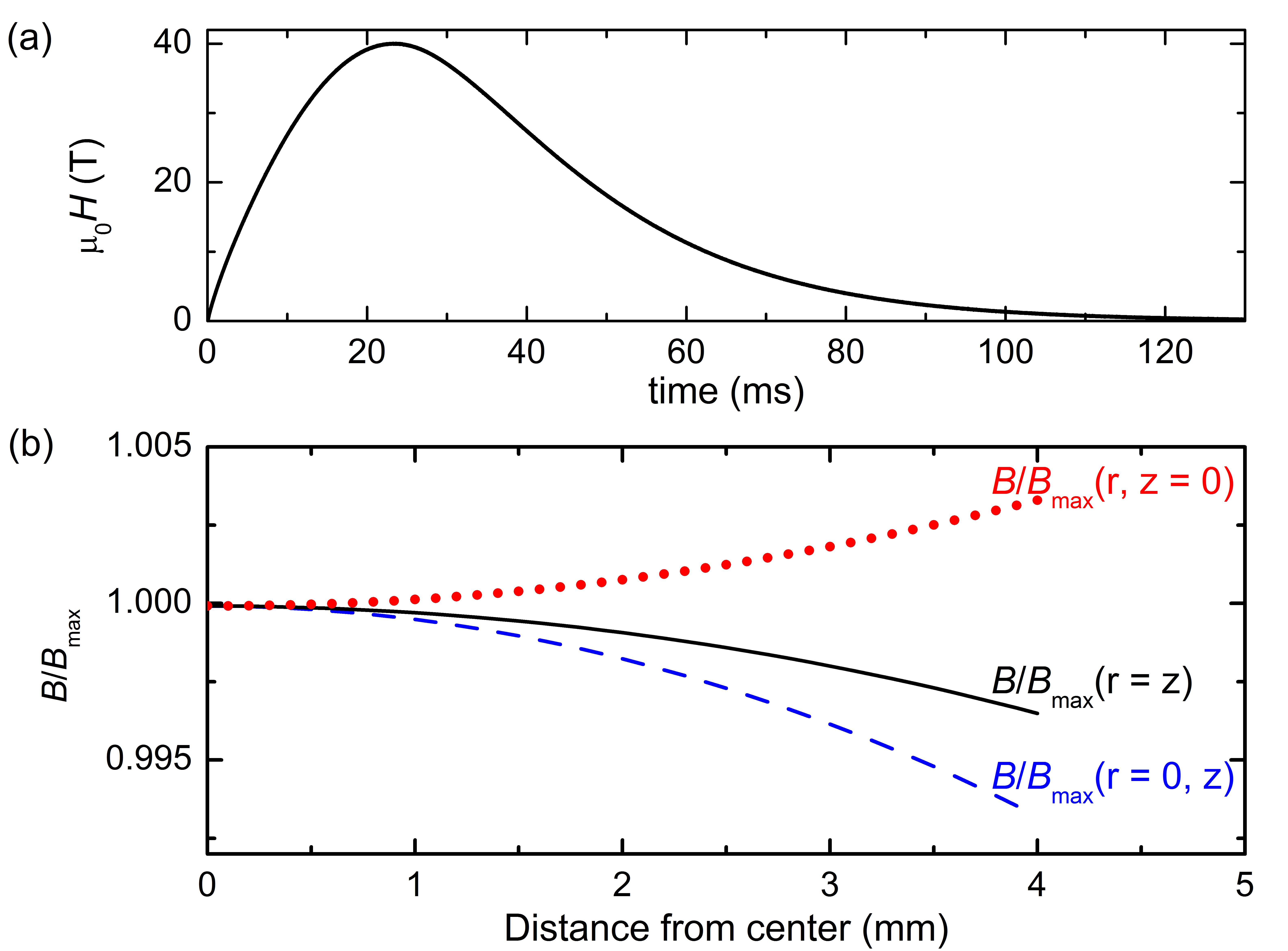}
\caption{(a) Time profile of the field pulsed up to 40\,T.
(b) Magnetic field homogeneity calculated from the centre of the magnet.
The longitudinal (axial) and radial distribution are denoted by $B$(z) and $B$(r), respectively.
}
\label{Fig2}
\end{figure}

\begin{figure*}
\centering
\includegraphics[width=1.7\columnwidth]{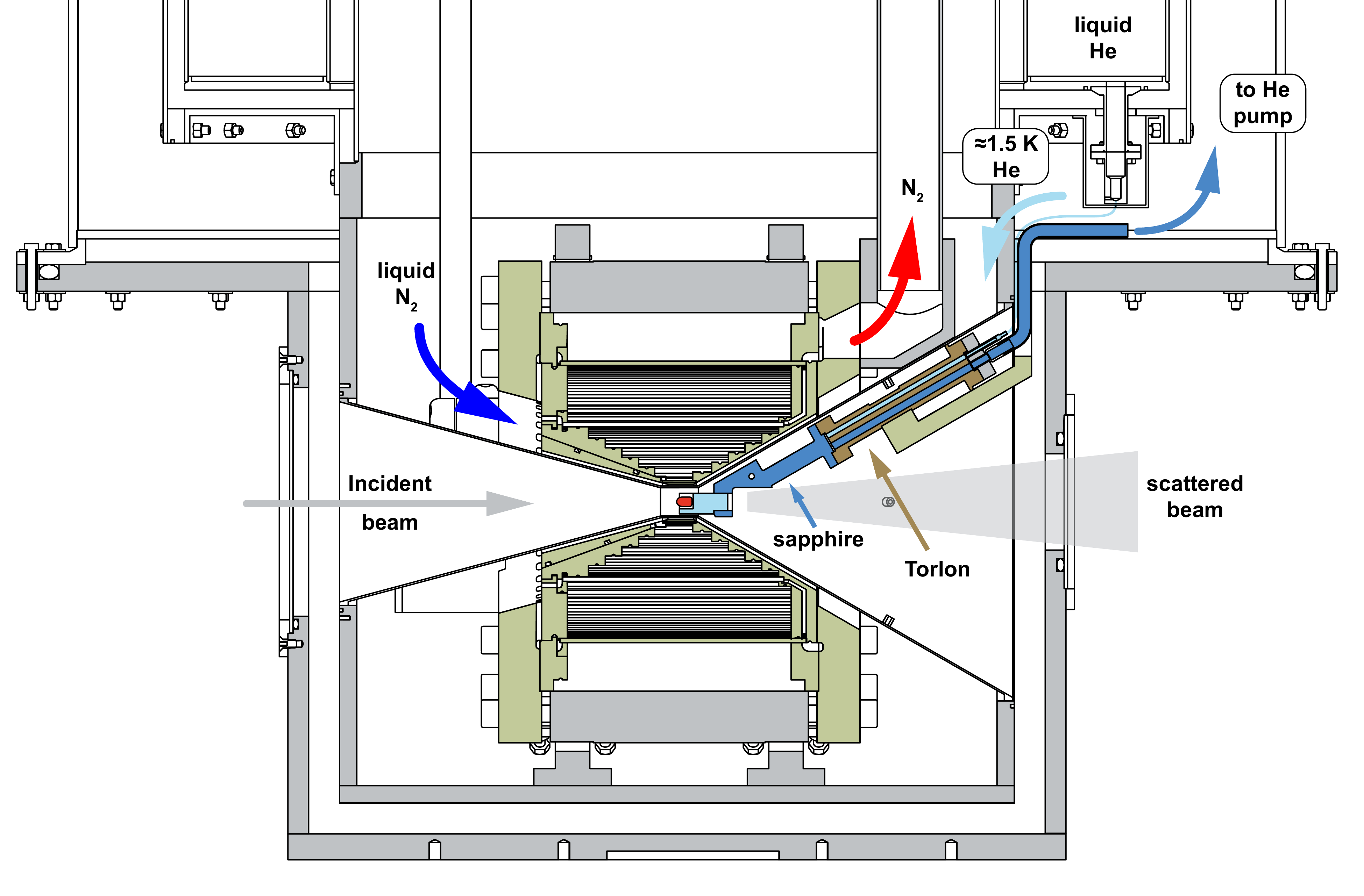}
\caption{Overview of the tail of the cryomagnet. The coil is immersed in liquid nitrogen and the gas evaporated after a pulse is evacuated through a chimney. The sample is under vacuum and fixed at the end of a cold finger made from sapphire. This finger is cooled down with a helium flow controlled with a cold-valve. The temperature is stabilised with a heater mounted on the finger.}
\label{cryomagnet}
\end{figure*}

The resulting magnet has an inductance of 60\,mH and a resistance of 259 and 1632 m$\Omega$
at liquid nitrogen and room temperature, respectively.
Using a 1.15\,MJ ($C=4$\,mF, $V=24$\,kV) transportable power supply \cite{current} developed at the LNCMI-Toulouse \cite{Duc2014},
it produces a maximum pulsed-field of 40\,T with a rise time of 23\,ms and a total
duration of more than 100\,ms every 9\,min (Fig. \ref{Fig2}(a)). At lower fields, the time between pulses
is reduced: every 7 and 5 min at 36 and 31\,T, respectively.
For fields below 20\,T, the duty cycle is limited by the time necessary to charge the capacitors, i.e., about a minute.
The duty cycle of this new cryomagnet is of 28 s per day at 30\,T,
and of 16 s per day at 40\,T \cite{Knafo2016}, with $t_{90\%}\simeq15.5$\,ms ($t_{90\%}$ is
defined as the time per pulse during which the field is greater than 90\% of the maximum field), meaning that
we have a field higher than 36\,T for 2.5 s per day.\\
These performances remain far more superior than those of mini-coil systems, which store
less energy but have shorter pulse duration.
Neutron scattering experiments \cite{Yoshii2009, Matsuda2010, Kuwahara2013}
performed with miniature coils producing field pulses of less than 10\,ms duration had indeed an
overall duty cycle of less than 2 s per day for a maximum field value of 30\,T, with $t_{90\%}\simeq1.9$\,ms.


At the centre of the magnet, the field factor $B/I$ is equal to $8.31\pm 0.02$ T/kA
(i.e. $I/B = 120.3\pm 0.3$ A/T) \cite{fieldfactor} and shows less than 1\% of variation
over a sphere of radius 4\,mm (see Fig.~\ref{Fig2}(b)).
This indicates that the magnetic field remains relatively constant over a large volume.

The coil is installed horizontally inside a cryostat developed by the ILL
and constructed by AS Scientific (UK). To allow easy replacement and efficient cooling
of the coil, the liquid nitrogen bath is central, large and surrounded by a liquid helium jacket
(see Fig.~\ref{coil_MAGFINS}). The initial cool-down time is 2.5 hours.
The maximum liquid nitrogen consumption to cool down the magnet is $\simeq 1 000$\,litres per day at 40\,T.
To cope with this demand, a vacuum insulated pipeline has been installed to automate the refills. It is connected to a 10.000 litres tanks shared with other instruments, which allows to perform experiments for several days.

The vertical access for the neutron beam is limited by the presence of the heat exchanger to $\pm 7^\circ$. This opening matches the height of the $\simeq 315\,$mm $\times 100$ mm window of the scattered beam. To minimise background issues, the incident and outgoing windows are both made from 2\,mm thick single crystalline silicon.

\section{Sample temperature control}
\begin{figure}
\includegraphics[width=0.95\columnwidth]{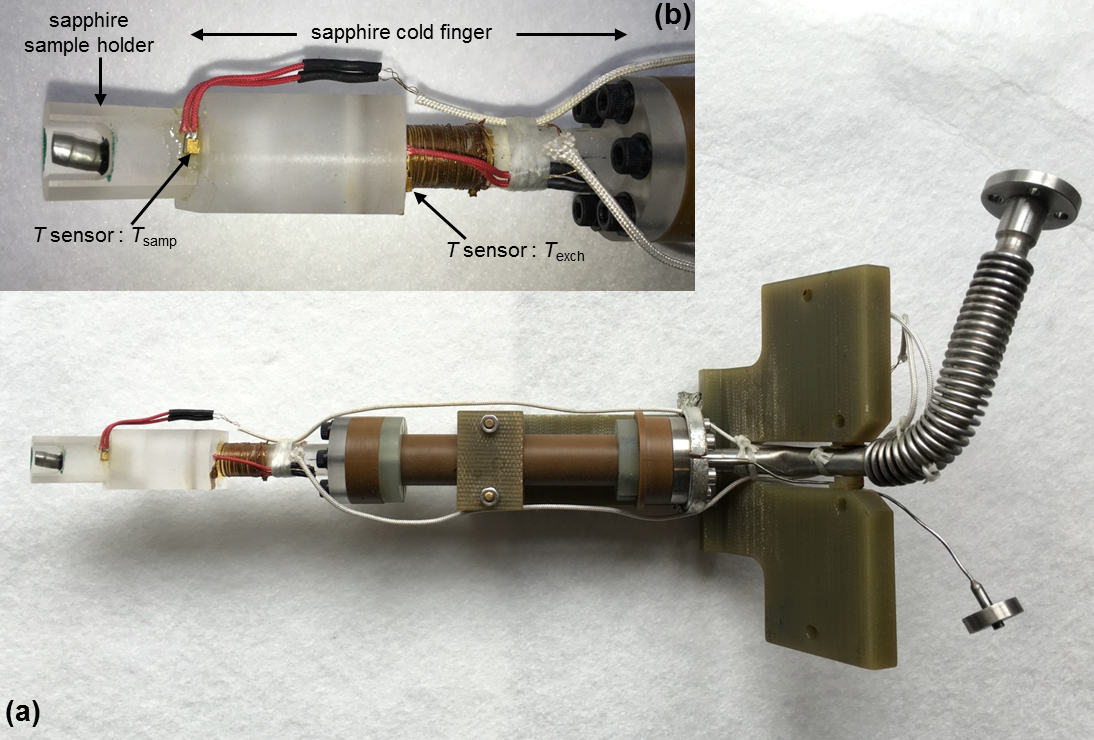}
\caption{Photos of the heat exchanger:
(a) the Joule-Thomson thermal expansion of $^4$He is performed against the flange of sapphire screwed at the end of the Torlon tube in which $^4$He circulates. Liquid $^4$He is injected into the capillary that can be seen at the right bottom of the image. The flexible pipe through which $^4$He is pumped is also shown. (b) Detailed assembly of the sapphire cold finger and sample holder.
}
\label{Sapphire}
\end{figure}	
The main current leads being fixed on one side of the coil, we have installed the heat exchanger on the opposite side along the top surface of the cone onto which the coil is wound (scattered beam side) (see Fig. \ref{cryomagnet}). In order to reach 2\,K, a standard ILL cold-valve has been installed at the bottom of the liquid $^4$He bath on the same side to reduce the length of the capillary driving liquid $^4$He to the heat exchanger. The Joule-Thomson thermal expansion is performed by pumping $^4$He in the heat exchanger with a pump connected to a \O30\,mm tube located at the top of the cryostat. The sample is placed in the central \O16.5\,mm cylinder where the magnetic field is maximal without touching the double-cone cooled down to liquid nitrogen temperature.

To avoid eddy currents and minimise neutron background, the heat exchanger (developed at the ILL) was made in Torlon and sapphire (Fig.~\ref{Sapphire}). Torlon is insensitive to magnetic fields and rigid enough to build flanges tight to $^4$He. Two holes were drilled into the Torlon bar: one of \O3\,mm for injecting the liquid $^4$He coming from the cold-valve and one of \O4.5\,mm for pumping it out. The injection is performed with a \O0.6$\times$1\,mm capillary brazed to the cold-valve flange and inserted in the \O3\,mm hole up to its end. The thermal expansion is thus realised at the end of the Torlon bar against a piece of single crystalline sapphire called cold-finger and shown in Figure~\ref{Sapphire}(b). It consists of a bar wound with a Manganin wire for regulating the temperature (heater) and an octagonal hole in which the sample holder is placed. This shape prevents the sample from rotating around the axis of the coil. With the provided sample holders, it also allows a quick rotation of the single crystalline sample by $30^\circ$ or $45^\circ$ without realignment and regluing \cite{rotation}.

The length and shape of the cold finger was designed to reduce as much as possible the presence of sapphire in the beam and to avoid neutron scattering by the Torlon bar. Two thermometers are installed: one near the heater ($T_{exch}$) and one on the sample holder ($T_{samp}$). To prevent heat production by eddy currents, we have chosen to glue compact Cernox CX-1050-SD sensors packed inside a sapphire base with alumina body and lid\cite{LakeShore}.

The sample is glued on a holder also made from single crystalline sapphire and sits in the vacuum of the cryostat. The maximum available volume is $8\times6\times6$\,mm$^3$.
The incident monochromatic beam hits the sample and then crosses 24\,mm of sapphire as shown in Figure~\ref{Sapphire}.
The transmission is about 96\% at $\lambda_i=2.36$\,\AA\ (or $k_i=2.662$\,\AA$^{-1}$) and almost 98\% at $\lambda_i=1.52$\,\AA\ (or $k_i=4.1$\,\AA$^{-1}$).

The base temperature measured near the sample is slightly less than 2\ K. This is a little more than the 1.5\,K expected with a helium flow system because of the radiation of the surrounding screens thermalised at 77\,K. The sample temperature can be stabilised at any value up to room temperature and the cool down time does not exceed 30\,min. The helium boil-off rate is of $\approx1$\,litre/hour at base temperature which leads to an autonomy of 1.5\,day.

\section{Application to neutron diffraction}
\subsection{Data acquisition scheme and analysis}
\begin{figure}
\includegraphics[width=\columnwidth]{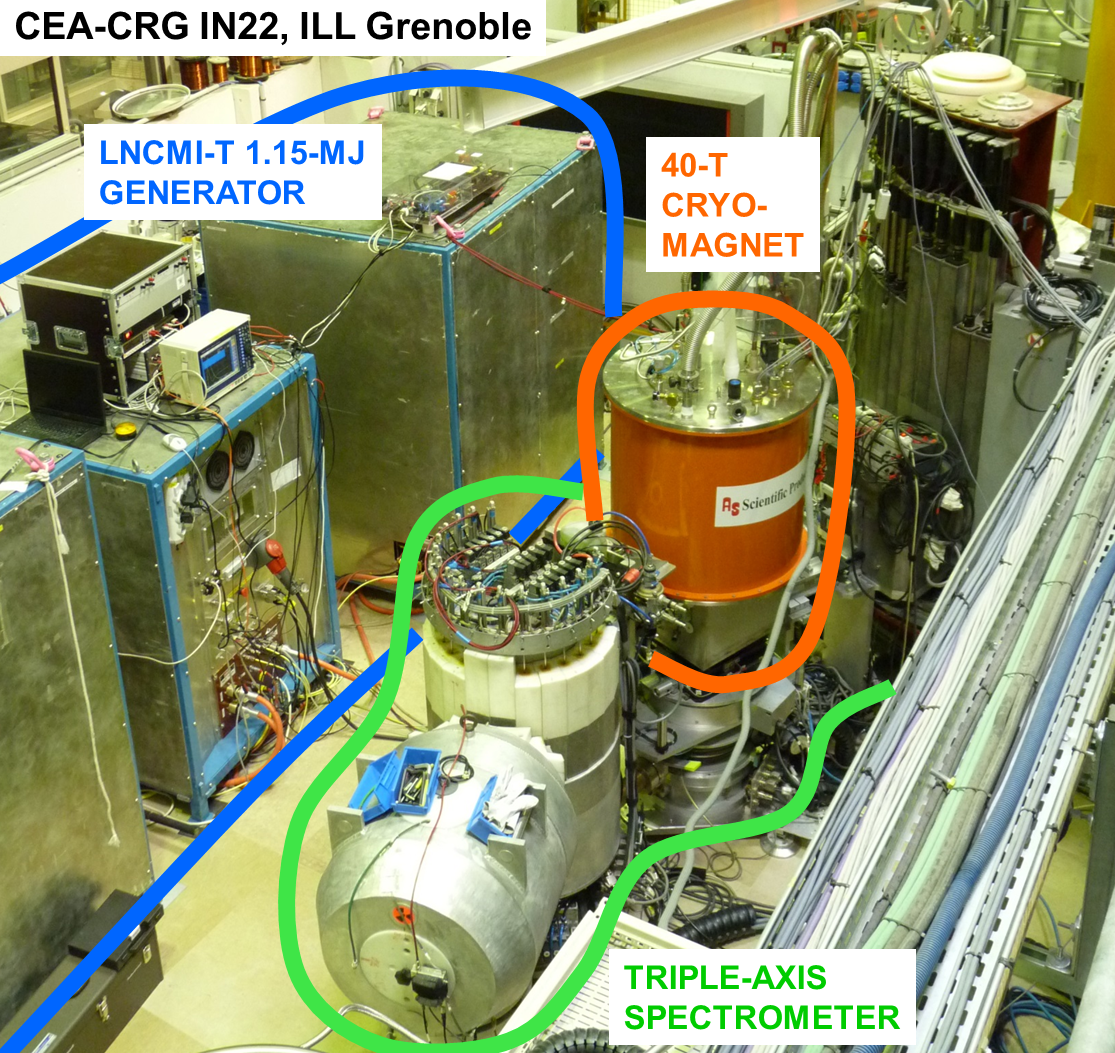}
\caption{IN22 triple-axis spectrometer equipped with the mobile pulsed magnet system consisting
of the mobile 1.15\,MJ pulsed field generator and the 40-tesla cryomagnet.}
\label{SetupIN22}
\end{figure}	
\begin{figure}
\includegraphics[width=\columnwidth]{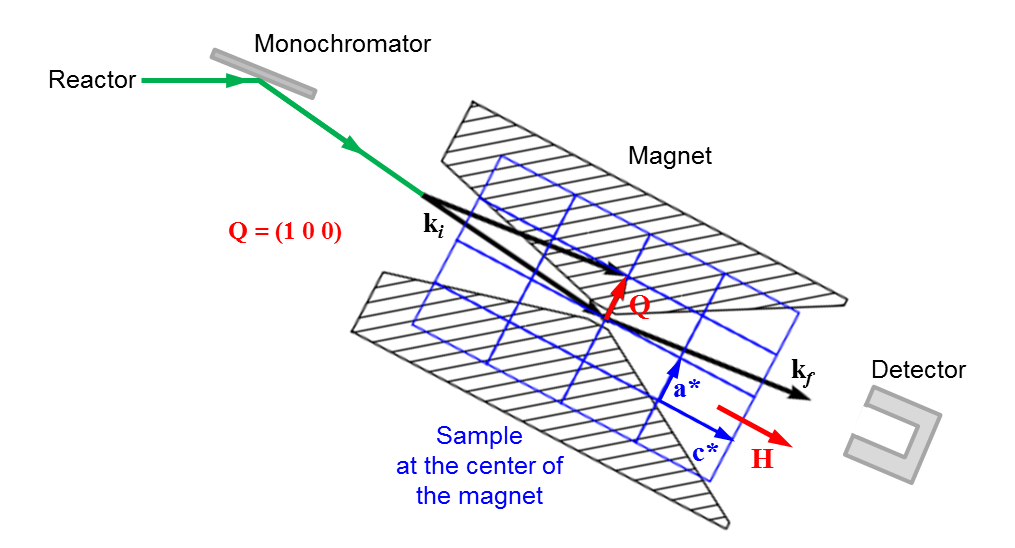}
\caption{Schematic view of the experimental set-up installed on the triple-axis spectrometer IN22 operated in double-axis mode. An example of neutron scattering geometry in the environment of the 40-tesla cryomagnet is shown. The direction of the incident and scattered beams are given by the wavevectors \textbf{k$_i$} and \textbf{k$_f$}, respectively. The reciprocal lattice of the sample located at the centre of the magnet is represented by the blue lattice. The scattering vector \textbf{Q}=\textbf{k$_i$}-\textbf{k$_f$} and the direction of the applied magnetic field are indicated in red.}
\label{Setup}
\end{figure}	

The efficiency of this new 40-tesla cryomagnet was tested and commissioned on the IN22 spectrometer at the ILL in elastic mode without any modification of the spectrometer. The coil was energised by the transportable 1.15\,MJ pulsed field generator of the LNCMI Toulouse (Fig.~\ref{SetupIN22}). A schematic view of the neutron scattering geometry is shown Figure \ref{Setup}.

The IN22 spectrometer is operated either in double-axis or triple-axis mode depending on the sample used. In this later mode, the signal-to-noise ratio can be improved by using a pyrolytic graphite (PG) analyser crystal. The incident neutron
wavelength is chosen between $\lambda_i=1$\,\AA\ and $\lambda_i=2.36$\,\AA\ with a PG monochromator in order to avoid the blind zone of the magnet. The sample is usually oriented with a crystallographic axis parallel or slightly misaligned with the horizontal magnetic field to reach the most interesting part of the reciprocal space.

Neutron counts are measured with a $^3$He fast single-detector starting saturating above 18000 neutrons per second. The detector is oriented horizontally in the double-axis mode but vertically in the triple-axis mode to take advantage of the horizontal focalisation. Neutron pulses are recorded with a digital data recorder (HIOKI Memory HiCorder model 8860) measuring the generator voltage and current, and the voltage on a pick-up coil. After correction for the neutron time of flight, the field dependence of the neutron intensities at each momentum transfer is extracted by summing data accumulated over a few identical pulsed-field shots, with either constant time- or constant field-integration windows (rises and falls of the pulsed field are analysed sepa\-rately).

\subsection{Application to the frustrated antiferromagnet T$\textit{b}$B$_4$}
\begin{figure}
\includegraphics[width=\columnwidth]{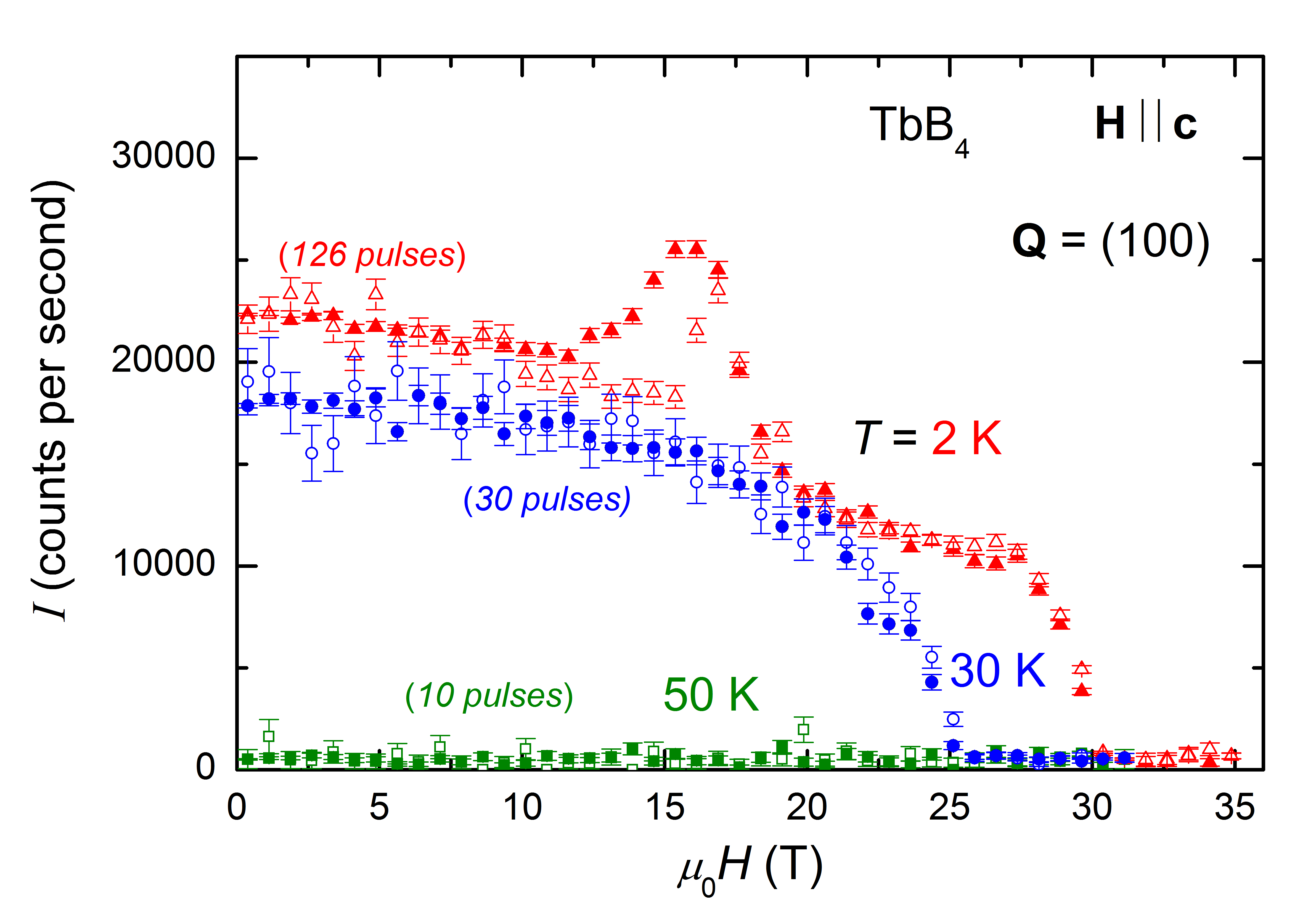}
\caption{Field dependence of the diffracted intensity at the momentum transfer \textbf{Q}~=~(1~0~0) in fields up to 35\,T at \textit{T}~=~2, 30 and 50\,K, measured with the 40-tesla cryomagnet. The magnetic field was applied along the \textit{c}-axis. Data are integrated within field steps $\Delta(\mu_0$\textit{H})= 0.75\,T. The error bars $\Delta$\textit{I} are given by the square root of the neutron counts ($\Delta$\textit{I}= $\sqrt{I/\tau}$ where $\tau$ is the total
accumulation time over the number of pulses indicated for each temperature). Open and full symbols correspond to rising and falling fields, respectively.}
\label{Tdep}
\end{figure}	

\begin{figure*}
\includegraphics[width=1.9\columnwidth]{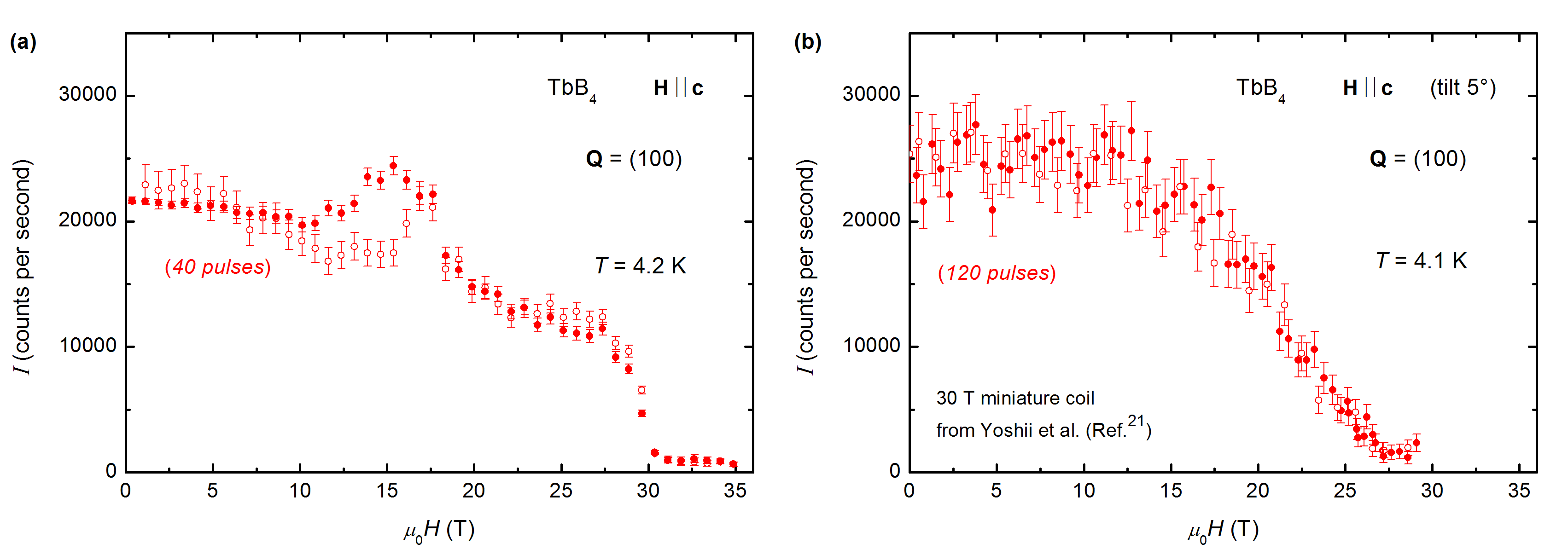}
\caption{(a) Field dependence of the diffracted intensity at the momentum transfer \textbf{Q} = (1 0 0) in fields up to 35\,T at \textit{T}~=~4.2\,K, measured with the new cryomagnet. The magnetic field was applied along the \textit{c}-axis. Data are integrated within field steps
$\Delta(\mu_0$\textit{H})= 0.75\,T. The error bars $\Delta$\textit{I} are given by the square root of the neutron counts
($\Delta$\textit{I}= $\sqrt{I/\tau}$ where $\tau$ is the total accumulation time over the number of pulses indicated for each temperature).
(b) Field dependence of the diffracted intensity at the same momentum transfer in fields up to 30\,T
measured with the miniature coil system (Ref.~\cite{Yoshii2009}). The magnetic field was applied horizontally with a tilt of $5^\circ$ from the $c$-axis of the crystal. Neutron counts are summed over 120 identical pulsed-field shots, with constant time-integration windows $\Delta\textit{t}=$ 40\,$\mu$s. For both data sets, open and full symbols correspond to rising and falling fields, respectively.}
\label{TbB4}
\end{figure*}	
To evaluate the performances of this coil and setup, the frustrated antiferromagnet TbB$_4$ was reinvestigated.
In this system, two successive antiferromagnetic (AF) transitions occur at
$T_{N1} \simeq$ 44\,K and $T_{N2} \simeq$ 24\,K \cite{Fisk1981},
leading to a quasi-planar non-collinear AF structure described by the propagation vector \textbf{k}~=~(0~0~0) \cite{Matsumura2007}.
Upon the application of a magnetic field along the \textit{c}-axis, TbB$_4$ is known for its devil-staircase-like magnetization process,
stabilising successive magnetization plateaux for fields $\mu_0H>16$\,T \cite{Yoshii2008}.
Because of the high magnetic moment of Tb$^{3+}$ ions, TbB$_4$ has been one of the first compounds studied by neutron diffraction
combined with pulsed field miniature coils \cite{Yoshii2009}.
A model of magnetic structure consisting of Ising spins and XY spins has been proposed to explain
the half-magnetization state \cite{Yoshii2009, Inami2009}.

For the present work, single crystals of TbB$_4$ enriched with $^{11}$B were kindly provided by Iga
(Ibaraki University) and Michimura (Saitama University).
A single crystal ($\sim2\times2\times1$ mm$^3$, $m \simeq 30$\,mg) was mounted
on the sapphire sample holder with the scattering plane defined by [1~0~0] and [0~0~1].
Magnetic field was applied along the $c$-axis.
In Figure~\ref{Tdep}, the field dependence of the diffracted intensity measured at the momentum transfer
\textbf{Q}~=~(1~0~0) in fields up to 35\,T and at tempera\-tures \textit{T}~=~2, 30 and 50\,K
is reported. At $T=$~2\,K, these data
exhibit stepwise variations of the diffracted intensity, highlighting at least three transitions and three AF phases.
When the magnetic field is applied, the intensity shows a step-like increase at $\mu_0H=16/14$\,T (rising/falling field)
followed by a step-like decrease at $\mu_0H\simeq18$\,T
and finally vani\-shes at $\mu_0H=30$\,T.
At $T=$ 30\,K $> T_{N2}$, the intensity decreases continuously up to $\mu_0H_c=$ 25\,T, indicating
a single AF phase in fields up to $\mu_0H_c$, in agreement with the $H$-$T$ phase diagram determined by magnetization
measurements \cite{Yoshii2008}.
At $T=$ 50\,K~$> T_{N1}$, the intensity is at the background level, attesting the absence of magnetic diffraction.

In Figures~\ref{TbB4}(a) and ~\ref{TbB4}(b), a comparison of the field dependences of the diffracted intensity measured on TbB$_4$
at $T=4.2$\,K and \textbf{Q}~=~(1~0~0) using the new cryomagnet and the mini-coil system \cite{Yoshii2009}, respectively,
illustrates the improved sensitivity of our cryomagnet to detect fine changes of magnetic structures.
We note that, with the mini-coil system, the magnetic field was applied horizontally with a tilt of $5^\circ$ from the $c$-axis
(dimensions and weight of the single crystal used in Ref.~\cite{Yoshii2009} are $4\times4\times3$\,mm$^3$ and $m\simeq296$\,mg).
Similar features to the ones measured at $T=2$\,K are detected at $T=4.2$\,K with the new 40-tesla cryomagnet
whereas they were not observed in the previous experiment.
We note that even with a smaller number of pulses, the new data show better counting statistics.

The detailed study of TbB$_4$ by neutron diffraction under pulsed magnetic fields using our new cryomagnet will be reported in a forthcoming paper.

\section{Conclusion}

We have developed a 40-tesla pulsed-field cryomagnet for neutron diffraction featuring
an unprecedented duty cycle of 28 s per day at 30\,T and 16 s per day at 40\,T.
The sample temperature is regulated between 2 and 300\,K independently from the pulse fields generation.
We illustrate the performances of this magnet with the reinvestigation of the high
field magnetic structures of the frustrated antiferromagnet TbB$_4$.
At low temperature, the field dependence of the diffracted intensity
of the antiferromagnetic reflection (1~0~0) shows step-like variations
which reveal subtle changes in the magnetic structures of the different magnetization plateau phases.
These new features were observed thanks to higher counting statistics and
a precise control of the sample temperature resulting both from the combination of a longer
pulse duration, a higher duty cycle and a longer rising time (much less heating by eddy currents)
than mini-coil systems.

More than four thousand field pulses, with 65\% of them at more than 30\,T, have
already been generated with this magnet. They have allowed us to investigate various magnetic systems such
as heavy fermions materials \cite{Knafo2016, Kumar} and quantum spins systems \cite{Gazizulina}.

This equipment is available to the ILL users through a scientific collaboration with the LNCMI.

Several improvements are planned for the near future. First we will implement a high
counting rate, high efficiency position sensitive detector, having good-enough spatial
and time resolutions. Then, modifications of the cryogenic environment will be made, with the target to extend
the temperature range down to 1.4 K.

\begin{acknowledgments}
The authors are very grateful to F. Lecouturier and N. Ferreira
who performed the characterization of the wire conductor (resistivity and wire
strength measurements) and to J.-M. Lagarrigue who machined several pieces required for
the building of the coil.
The authors would like to thank H. Nojiri for very helpful technical and scientific discussions.
The CEA-CRG Grenoble and the ILL are greatly acknowledged
for granting the beam time for these developments and experiments.
This work was financially supported by the French National Research Agency
(ANR project MAGFINS: Grant N$\circ$. ANR-10-BLN-0431) and by
the program Investissements d'Avenir ANR-11-IDEX-0002-02 (reference ANR-10-LABX-0037-NEXT).

\end{acknowledgments}


\begin{thebibliography}{39}
\bibitem [{dut()}]{duty}%
  \BibitemOpen
  \href@noop {} {}\bibinfo {note} {Defining $t_{useful}$ as the useful time of
  a field pulse and $t_{cool}$ as the time to cool down the magnet between two
  pulses, the duty-cycle of a pulsed field coil is defined by the ratio
  $t_{useful}/t_{cool}$. In the neutron experiments described here, data are
  recorded over the whole pulse. So, $t_{useful}$ corresponds to the total
  pulse duration.}\BibitemShut {Stop}%
\bibitem [{rot()}]{rotation}%
  \BibitemOpen
  \href@noop {} {}\bibinfo {note} {The rotation of the sample is only possible
  at room temperature. To do this handling, the liquid nitrogen is removed and
  the cryomagnet is warmed up to room temperature before breaking the vacuum.
  The cryomagnet is then opened to have access to the cold finger and sample
  holder.}\BibitemShut {Stop}%
\end{thebibliography}
%

\end{document}